\documentclass[a4paper,twocolumn,11pt]{quantumarticle}
\pdfoutput=1
\usepackage{amsmath}
\usepackage{amssymb}
\usepackage{graphicx}
\usepackage{hyperref,cleveref}
\usepackage{url}
\usepackage{float}

\makeatletter
\usepackage{etoolbox}
\patchcmd\H@refstepcounter{\protected@edef}{\protected@xdef}{}{}
\makeatother

\begin{document}

%opening
\title{A Dyson equation approach for averaging of classical and quantum observables on multiple realizations of Markov processes}
\author{Simone Sturniolo}
\affiliation{Scientific Computing Department, UK Research and Innovation, Harwell Campus, Didcot OX11 0GD (UK)
}
\email{simone.sturniolo@stfc.ac.uk}
%\pacs{02.50.Ga, 76.60.−k, 76.75.+i}

\begin{abstract}	
	Time dependent signals in experimental techniques such as Nuclear Magnetic Resonance (NMR) and Muon Spin Relaxation ($\mu$SR) are often the result of an ensemble average over many microscopical dynamical processes. While there are a number of functions used to fit these signals, they are often valid only in specific regimes, and almost never properly describe the ``spectral diffusion'' regime, in which the dynamics happen on time scales comparable to the characteristic frequencies of the system. Full treatment of these problems would require one to carry out a path integral over all possible realizations of the dynamics of the time dependent Hamiltonian.\newline
	In this paper we present a numerical approach that can potentially be used to solve such time evolution problems, and we benchmark it against a Monte Carlo simulations of the same problems. The approach can be used for any sort of dynamics, but is especially powerful for those that can be approximated as Markov processes, in which the state at each step only depends on the previous state of the system. The approach is used to average both classical and quantum observables; in the latter case, a formalism making use of Liouvillians and density matrices is used.
\end{abstract}

\maketitle

\section{Introduction}

The time evolution of open quantum systems is a well known problem in quantum dynamics, with key implications for a variety of fields. Many approaches have been developed to deal with it, in presence of both Markovian and non-Markovian dynamics, such as the Lindblad master equation \cite{breuer2003, manzano2020}.\newline
In the field of spin resonance techniques, such as Nuclear Magnetic Resonance (NMR) and Muon Spin Relaxation ($\mu$SR), the theory of open quantum systems plays a role in the interpretation of many results when either the dynamics of the magnetic fields or of the atomic nuclei themselves introduce a time dependence in the Hamiltonian. Such dynamical processes play an important role in loss of coherence and thus apparent decay of magnetisation, a key phenomenon in both solid state NMR \cite{abragam1961, slichter1990} and $\mu$SR \cite{reotier1997,blundell1999}. In these fields it is common to treat such problems with approximations that work only when the dynamical processes are far removed from the time scales on which the system's evolution takes place, either by being much slower or much faster. The intermediate regime, known as ``spectral diffusion'' \cite{bloembergen1948, kindel2014}, can not in general be described with a simple closed form expression.\newline
In a previous paper \cite{sturniolo2012}, this problem was addressed for a particular case, two dipolar coupled spins at a fixed distance rotating freely in the spherical angle. Since the solution of the quantum problem for that case is known, the problem was reduced to a classical one. A distribution of possible frequencies $\omega(\theta, \phi)$ was established on the solid angle, and given diffusive dynamics on the sphere, the signal was described as a sum over all possible dynamical paths

\begin{equation} \label{fid_path}
\overline{S}(t) = \mathrm{Re} \int d\psi(t) p[\psi(t)]e^{i\int_0^t\omega(\tau)d\tau},
\end{equation}

where we intend the $\psi(t)$ to be all possible trajectories in time followed by the system, the functional $p[\psi(t)]$ the probability weighing each trajectories, embodying the properties of the dynamics, and $\omega(\tau)$ the precessing frequency at a given time. This integral obviously reverts to a much simpler one in the case in which $\omega$ has no time dependence. An analytical solution was then provided for the special case described, of dipoles rotating on a sphere. However it must be noted that \cref{fid_path} describes the entire class of problems in which an ensemble of systems accumulate a phase through a time-dependent frequency $\omega(t)$. The only difference lies in how the various possible paths are weighed.\newline
Let us now consider a different problem - an ensemble of classical vectors $\mathbf{v}$ rotating in space around an axis, and at an angular speed, that are both time dependent and subject to some kind of random dynamics. This could for example be a semiclassical description of a single spin; the precession dynamics of a spinning body under a variable force. In this case, the average vector observed as a function of time could be written as

\begin{equation}\label{vector_path}
\overline{\mathbf{v}}(t) = \int d\psi(t)p[\psi(t)]e^{\int_0^t\omega(\tau)\mathbf{a}(\tau)\mathbf{L}d\tau} \mathbf{v},
\end{equation}

where $\mathbf{a}(t)$ is the time dependent axis of rotation and $\mathbf{L}$ the vector of the infinitesimal generators of the Lie group of rotations \cite{rossmann2006}. In this case, the exponential is a \textit{matrix} exponential; \cref{vector_path} works in the approximation that $e^Xe^Y = e^Z = e^{X+Y}$ for matrices even if X and Y do not commute. This is not true in general; however, when computing numerically the integral in \cref{vector_path} in practice, we discretize time in small intervals of length $\delta$. This means the matrices X and Y will both be of the order of the chosen time step, and, as one can see from the Baker-Campbell-Hausdorff formula \cite{rossmann2006}, in this case,

\begin{align}\label{baker_campbell}
Z &= X + Y + \frac{1}{2}[X, Y] + \frac{1}{12}[X,[X,Y]] -  \frac{1}{12}[Y,[X,Y]] + ... \nonumber \\
&= X + Y + \mathcal{O}(\delta^2).
\end{align}

Therefore, the approximation should be satisfactory provided that the time step is chosen to be small enough, and becomes exact in the limit of an infinitesimal time step.\newline
Finally, let us consider the case of an ensemble of quantum systems, starting in a pure state represented by the density matrix $\rho_0$. It is usual to write the evolution of the density matrix using its commutator with the Hamiltonian; here, however, it is more convenient to work in the so-called Fock-Liouville space, in which the density matrix can be written as a vector, $\rho_0 \rightarrow \left|\left.\rho_0 \right>\right>$, and the Liouvillian superoperator as a matrix $\tilde{\mathcal{L}}$ \cite{manzano2020}. Then the evolution in presence of a randomly varying Hamiltonian (and thus, Liouvillian), can be written as

\begin{equation}\label{quantum_path}
\left|\left.\rho \right>\right> (t) = \int d\psi(t)p[\psi(t)]e^{\int_0^t\tilde{\mathcal{L}}(\tau)d\tau} \left|\left.\rho_0 \right>\right>.
\end{equation} 

The parallels between \cref{fid_path}, \cref{vector_path} and \cref{quantum_path} are obvious; they all represent a generalisation of the problem discussed in \cite{sturniolo2012}, but cover a wide range of potential problems in classical and quantum mechanics, and figure often in quantum field theory \cite{creutz1981,gies2002}. Here we set out to outline a general and simple method to numerically solve this class of problems; we focus in particular on the case of Markovian dynamics, for which the method is especially fast and powerful.

\section{Theory}

\subsection{General case}

In \cite{sturniolo2012}, the authors present an analytical formula to solve the path integral in \cref{fid_path} for the specific case of dipole pairs of fixed distance undergoing uniform rotational diffusion. Here we generalise that result to potentially any distribution of fields and correlation functions, as well as matrix exponentials.\newline
The first step is to derive a Dyson equation for the Green's function of the system. We define $\Omega$ as the entire domain of frequencies and $G_0[\omega_j, t; \omega_i, 0]$ as the free Green's function describing the dynamics of the system, expressing the probability for a spin starting from the site with frequency $\omega_i$ to end up in $\omega_j$ after a time $t$. We call $\psi(\omega_j, t; \omega_i, 0)$ a generic path the system can take such that it starts at time $0$ in the state $\omega_i$ and ends at time $t$ in the state $\omega_j$ (from here onwards, the dependencies are omitted and we will only write $\psi$). We also label as $p[\psi]$ the probability that a specific path will be realised under the system's dynamics. From this definition it also follows that 

\begin{equation}
G_0[\omega_j, t; \omega_i, 0] = \int_{(\omega_j, t; \omega_i, 0)} d\psi p[\psi],
\end{equation}

namely, that the total probability of all possible paths for given initial and final times and states $(\omega_j, t; \omega_i, 0)$ equals the free Green's function for those start and end points. Then the Green's function that we are looking for is:

\begin{widetext}
\begin{align}\label{dyson_eq_deriv}
G&[\omega_j, t; \omega_i, 0] = \int_{(\omega_j, t; \omega_i, 0)} d\psi p[\psi] e^{i\int_0^t\omega(\tau)d\tau} = \nonumber \\
&= \int_{(\omega_j, t; \omega_i, 0)} d\psi p[\psi] \left[1+i\int_0^t\omega(\tau)d\tau+...\right] = \nonumber \\
&= G_0[\omega_j, t; \omega_i, 0] + i\int_\Omega d\omega_1 \int_0^t d\tau_1 G_0[\omega_j, t; \omega_1, \tau_1] \omega_1\biggl[ G_0[\omega_1, \tau_1; \omega_i, 0]  + \biggr. \nonumber \\
&\biggl.i\int_\Omega d\omega_2 \int_0^{\tau_1} d\tau_2  G_0[\omega_1, \tau_1; \omega_2, \tau_2]\omega_2 G_0[\omega_2, \tau_2; \omega_i, 0] d\tau +...\biggr],
\end{align}
\end{widetext}

found by expanding the exponential in Taylor series, then replacing the powers of integrals with multidimensional integrals, with the $n!$ factors due to the Taylor expansion being cancelled out by the $n!$ ways in which the time intervals for each term can be ordered. \Cref{dyson_eq_deriv} shows a clearly recursive structure; we can then write it in finite form as:

\begin{align}\label{dyson_eq}
G&[\omega_j, t; \omega_i, 0] = G_0[\omega_j, t; \omega_i, 0] + \nonumber \\ 
+ &i \int_0^t d\tau \int_\Omega  d\omega G_0[\omega_j, t; \omega, \tau]\omega G[\omega, \tau; \omega_i, 0].
\end{align}

It must be noted that then the function in \cref{fid_path} is in fact given by the double average of $G$ over the initial and final $\omega$.\newline
We now consider a discrete version of \cref{dyson_eq}. We do this by considering only a finite number of possible frequencies $\omega$ and discretizing the time integral into $n$ intervals of length $\delta$. If we do that we can represent the Green's functions as matrices, with the initial and final states corresponding to row and column indices, and the indices for $\omega$ updating the fastest. In general, for a system with N different frequencies we would have:

\begin{equation}\label{matrix_def}
\mathbf{G}^{(0)}_{kN+i, lN+j} = G_0[\omega_j, l\delta; \omega_i, k\delta]
\end{equation}

and a diagonal matrix for the frequencies (written using the Kronecker delta):

\begin{equation}\label{omega_mat}
\mathbf{\Omega}_{i,j} = \delta_{ij}\omega_i.
\end{equation}

Consider that for the cases of \cref{vector_path,quantum_path}, the $\omega_i$ are actually matrices, and thus $\mathbf{\Omega}$ is a rank 4 tensor. However, this in practice does not affect the general argument, and we will see later how to accommodate for it.\newline
Now we can rewrite and solve \cref{dyson_eq} as:

\begin{align}\label{dyson_mat}
\mathbf{G} &= \mathbf{G}^{(0)} + i\mathbf{G}^{(0)}\mathbf{\Omega}\mathbf{G}\delta
\implies \nonumber \\
\implies \mathbf{G} &= ( \mathbb{I} -i\mathbf{G}^{(0)}\mathbf{\Omega}\delta)^{-1}\mathbf{G}^{(0)}.
\end{align}

The calculation requires the inversion of an $nN\times nN$ complex upper triangular matrix. The fact that the terms in the lower triangle are always zero is due to the restrictions imposed by causality - as they would describe a probability of propagation backwards in time in the $G_0$. The desired signal can then be retrieved by taking a partial trace over the $\omega$ indices and using only the first row of the resulting $\overline{\mathbf{G}}$.\newline
\Cref{dyson_mat} is very general. $G_0$ can be built to incorporate information about any kind of dynamics, and any distribution of fields can be described with a sufficiently high $N$. Precision of integration in time can be increased as well by increasing $n$ and reducing $\delta$. However, the operation of matrix inversion places a fundamental limit on how much the size of the system can be increased, especially if the function is to be used for data fitting. Matrix inversion by Gaussian elimination scales like $\mathcal{O}(n^3)$ \cite{farebrother1988}; algorithms with slightly more favourable scaling exist, but it is generally an expensive operation. Fortunately, as it will be shown in the next part, we can noticeably simplify the problem in the case in which $G_0$ describes a Markov process.

\subsection{Markovian dynamics}

Let us consider now what we will call the `small' matrices $\mathbf{G}^{(0)}_{i,j}$ so defined:

\begin{equation}\label{smallmat}
\mathbf{G}^{(0)}_{i,j} = \mathbf{G}^{(0)}[j\delta; i\delta].
\end{equation}

In other words, these matrices are the $N\times N$ blocks that compose the overall matrix $\mathbf{G}^{(0)}$ and characterise the free Green's function from a time $i\delta$ to a time $j\delta$ for all possible initial and final $\omega$. We similarly define the `small' $\mathbf{G}_{i,j}$. The defining property of a Markov process is that its state at each step only depends on the state at the previous step and on some probability transition matrix $\mathbf{P}$, giving the likelihood of going from one state to another \cite{markov2006,kemeny1974}. For a continuous time Markov process, the transition matrix can be defined in terms of the transition rate matrix $\mathbf{Q}$ \cite{kolmogoroff1931, asmussen2003} and the time step, as $\mathbf{P} = e ^{\mathbf{Q}\delta}$ is the solution of the backwards Kolmogorov equations. The only requirements for the transition matrix is that it must have all non-negative coefficients, and that the coefficients on each row sum up to one. For the transition rate matrix, the requirement is that it is antisymmetric (to conserve the total probability). It follows immediately from our earlier definition of the free Green's function that for Markovian dynamics it must be

\begin{equation}\label{greenf_transmat}
\mathbf{G}^{(0)}_{i,j} = \begin{cases}
\mathbf{P}^{(j-i)} \qquad & j \geq i \\
0 \qquad & j < i 
\end{cases}
\end{equation}

and in particular, $\mathbf{G}^{(0)}_{0,1} = \mathbf{P}$. 
A very common example could be the transition matrix for discretized uncorrelated jumping dynamics between N states, with a time step $\delta t$ and jumping frequency $\nu$, for which the $\mathbf{G}^{(0)}_{i,j}$ can be defined as:

\begin{equation}\label{G0_uncorr}
\mathbf{G}^{(0)}_{i,j} = H(j-i)\left[e^{-\nu(j-i)\delta}\mathbb{I}+\frac{1}{N}(1-e^{-\nu(j-i)\delta})\right]
\end{equation}

where $H$ is the Heaviside step function (taken to be $H(0) = 1$ and embodying the causality condition).\newline
In general, whenever \cref{greenf_transmat} holds, it becomes possible to simplify the solution of \cref{dyson_mat}. One can write the integral in time explicitly in its discrete form:

\begin{equation}\label{dyson_smallmat}
\mathbf{G}_{0,j}=\mathbf{G}^{(0)}_{0,j}+i\delta\sum_{i=0}^j\mathbf{G}^{(0)}_{i,j}\mathbf{\Omega}\mathbf{G}_{0,i}
\end{equation}

where $\mathbf{\Omega}$ stands for a `small' version of the matrix seen before with the same name. Remembering that $\mathbf{G}_{j,j}=\mathbf{G}^{(0)}_{j,j}=\mathbb{I}$ (which is obvious if one considers their physical meaning) it follows:

\begin{align}
\mathbf{G}_{0,j}=&\mathbf{G}^{(0)}_{0,j}+i\delta\mathbf{\Omega}\mathbf{G}_{0,j}+i\delta\mathbf{G}^{(0)}_{0,1}\sum_{i=0}^{j-1}\mathbf{G}^{(0)}_{i,j-1}\mathbf{\Omega}\mathbf{G}_{0,i} \nonumber\\ =&\mathbf{G}^{(0)}_{0,j}+i\delta\mathbf{\Omega}\mathbf{G}_{0,j}+\mathbf{G}^{(0)}_{0,1}\left(\mathbf{G}_{0,j-1}-\mathbf{G}^{(0)}_{0,j-1}\right) \nonumber \\
=&i\delta\mathbf{\Omega}\mathbf{G}_{0,j}+\mathbf{G}^{(0)}_{0,1}\mathbf{G}_{0,j-1} \nonumber\\
\implies&\mathbf{G}_{0,j} = (\mathbb{I}-i\delta\mathbf{\Omega})^{-1}\mathbf{G}^{(0)}_{0,1}\mathbf{G}_{0,j-1} \label{dyson_iter_final}
\end{align}

which provides us with an iterative relation through which the Green's function can be calculated step by step.\newline
A variant of this algorithm can be found if we rewrite \cref{dyson_smallmat} making use of the trapezoidal integration rule instead:

\begin{align}
\mathbf{G}_{0,j}=&\mathbf{G}^{(0)}_{0,j}+i\frac{\delta}{2}\left(\mathbf{\Omega}\mathbf{G}_{0,j}+\mathbf{G}^{(0)}_{0,1}\mathbf{\Omega}\mathbf{G}_{0,j-1}\right)+ \nonumber \\
&\mathbf{G}^{(0)}_{0,1}\left(\mathbf{G}_{0,j-1}-\mathbf{G}^{(0)}_{0,j-1}\right) \nonumber \\
=&i\frac{\delta}{2}\left(\mathbf{\Omega}\mathbf{G}_{0,j}+\mathbf{G}^{(0)}_{0,1}\mathbf{\Omega}\mathbf{G}_{0,j-1}\right)+\mathbf{G}^{(0)}_{0,1}\mathbf{G}_{0,j-1} \nonumber \\
\implies&\mathbf{G}_{0,j} = (\mathbb{I}-i\frac{\delta}{2}\mathbf{\Omega})^{-1}\mathbf{G}^{(0)}_{0,1}(\mathbb{I}+i\frac{\delta}{2}\mathbf{\Omega})\mathbf{G}_{0,j-1} \label{dyson_iter_trapz_final}
\end{align}

which provides more stability at the cost of only one more matrix multiplication.\newline
An important doubt can arise concerning the inverse of matrices in \cref{dyson_iter_final} and \cref{dyson_iter_trapz_final}. While their meaning is obvious in the case of the $\omega_i$ being scalars, it is far less so if they are matrices, and $\mathbf{\Omega}$ a rank 4 tensor. However, as it turns out, this is not in fact an issue. The $\mathbf{\Omega}$ matrix is diagonal; the inverse of $\mathbf{X} = \mathbb{I}-i\delta\mathbf{\Omega}$ can be simply written as

\begin{equation}\label{omega_invers}
X^{-1}_{ij} = \delta_{ij}(1-i\delta\omega_i)^{-1}.
\end{equation}

\Cref{omega_invers} remains entirely valid when $\omega_i$ is a matrix as well; all we need to do is replace $1 \rightarrow \mathbb{I}$ and perform a matrix inversion in place of an ordinary scalar division.

\section{Methods}

In house developed software was used for all tests and is provided as supplementary material. The code was written in Python, making use of the libraries Numpy \cite{numpy} and Scipy \cite{scipy} for vectorisation, Numba \cite{numba} for performance optimisation, random number generation and linear algebra operations. Monte Carlo simulations were performed by sampling a number of trajectories built to obey the transition rules of the Markov chain; these used Gillespie's algorithm \cite{gillespie1977} in combination with the transition rate matrix $\mathbf{Q}$, sampling the time interval for each transition from a state $i$ from an exponential distribution by generating a uniformly distributed random number $r \in [0, 1)$ and choosing

\begin{equation}\label{markov_gill_dt}
\delta t = -\frac{\log(r)}{\sum_i^{n}Q_{ij}},
\end{equation}

then picking a new state $j'$ so that, using a second random number $r$,

\begin{equation}\label{markov_gill_newstate}
\sum_j^{j'-1} Q_{ij} \leq r < \sum_k^{j'} Q_{ij}.
\end{equation}

Once a sequence of states and transition times is established, the relevant integral is computed piecewise. This algorithm uses an adaptive time step; it was chosen as it best approximates the dynamics, and was considered the best way to provide a good reference with a straightforward implementation of the process we're trying to describe. Its performance in terms of speed can vary, as the algorithm becomes much more computationally intensive for higher jumping frequencies (and thus, number of steps), and is thus not here the main focus of the comparison.\newline
For Green's function simulations, once the time step $\delta$ was fixed, the transition matrix $\mathbf{P} = e^{\mathbf{Q}\delta}$ was used. These simulations were ran with 10,000 steps of 0.001 each unless specified otherwise; Monte Carlo simulations were averaged over 500 trajectories. All calculations were ran on a laptop with an Intel i7-5500U 2.40 GHz processor and 12 GB RAM.

\section{Examples}

\subsection{Scalars}\label{subsec:scalar}

The first test we show here is a very simple application of \cref{fid_path}. In this simulation we consider the superposition of three frequencies $\omega_i$ which we've set to $[-1/2, 1, 2]$, in arbitrary units. We can consider these as the frequencies at three sites between which precessing spins can jump; this resembles, for example, the effect of chemical shifts of protons in freely rotating CH\textsubscript{3} groups \cite{hilt1964,wind1974}. Exchange between the three sites is free and determined, for a given frequency $\nu$, by \cref{G0_uncorr}. Clearly, the $\nu = 0$ limit for the observed signal is

\begin{equation}\label{slow_3sites}
\overline{S}_{slow}(t) = \frac{1}{3} \sum_{i=1}^3 \cos(\omega_it)
\end{equation}

whereas the $\nu \rightarrow \infty$ one is

\begin{equation}\label{fast_3sites}
\overline{S}_{fast}(t) = \cos(\bar{\omega} t),
\end{equation}

with the average frequency $\bar{\omega} = 1/2$ in this case.\newline

\begin{figure*}[hbtp]
	\includegraphics[width=\textwidth]{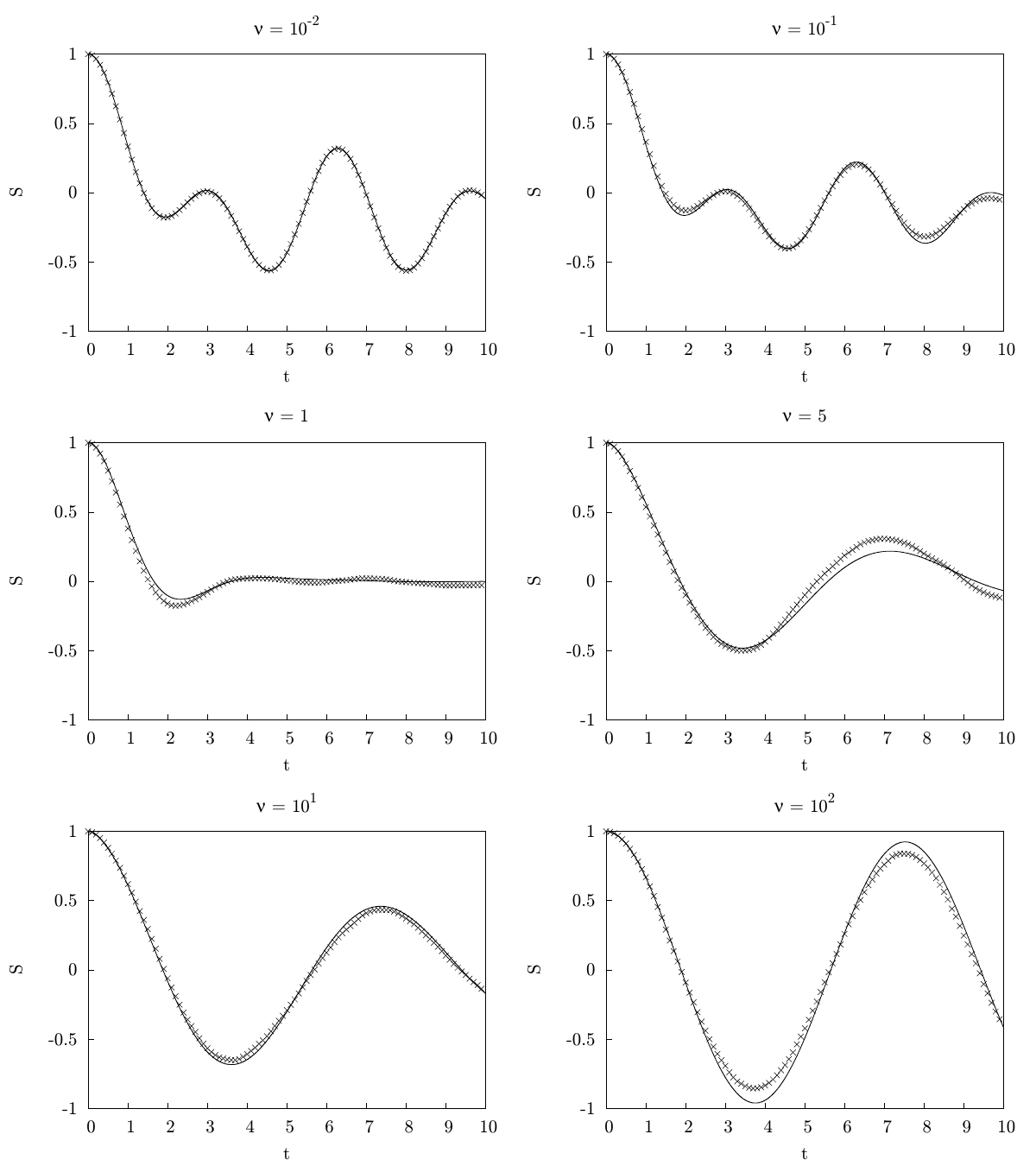}
	\caption{Total $S(t)$ for three overlapping sites with uncorrelated cross-site jumping for various frequencies $\nu$, with averaging over 100 Monte Carlo trajectories (crosses) and Dyson equation integrated using \cref{dyson_iter_trapz_final} (lines).}
	\label{fig:scalar}
\end{figure*}

\Cref{fig:scalar} shows the results for a range of intermediate frequencies. 
%Each of the Monte Carlo calculations took an average of 550 ms, versus 120 ms for the Dyson equation ones.
One can see both how the exchange affects the signal, by smoothing it down the most when $\nu$ is in the same order of magnitude as the frequencies $\omega_i$, before converging to a single cosinusoid at the average frequency $\bar{\omega}$. The Monte Carlo trajectories here and in the following examples are intentionally meant only for a qualitative comparison. Proper benchmarking against a high quality reference Monte Carlo calculation is provided in \Cref{subsec:performance}.

\subsection{Vectors}\label{subsec:vector}

\begin{figure}[h]
	\begin{center}
	\includegraphics[width=0.5\textwidth]{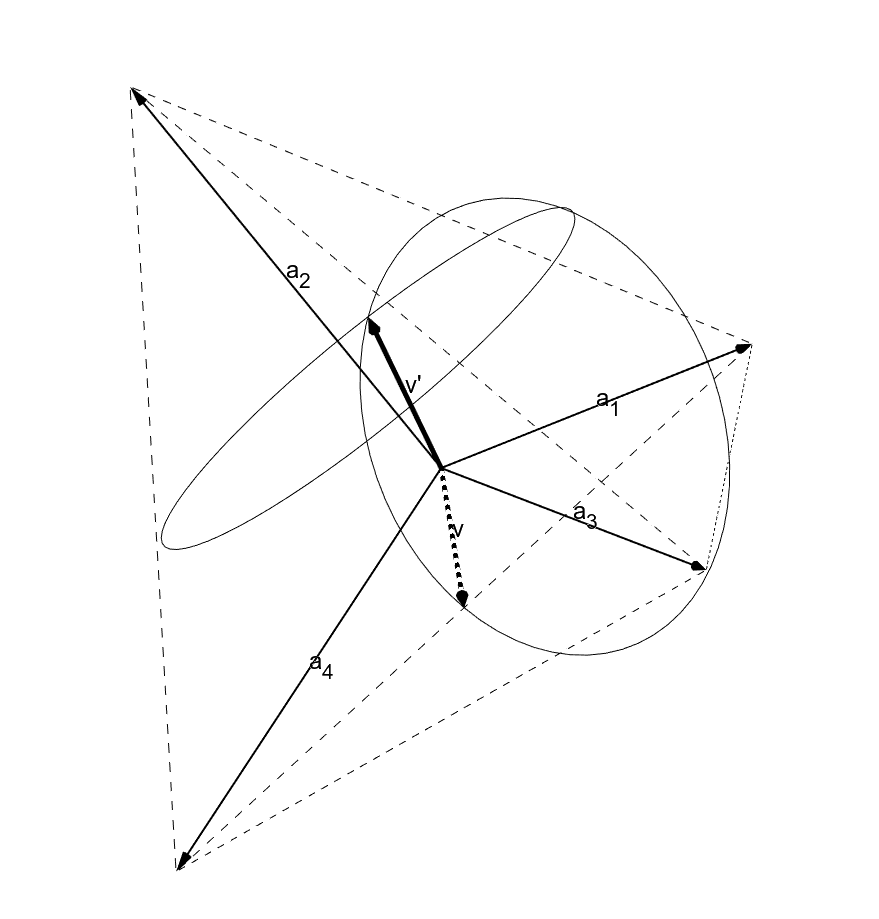}
	\caption{A schematic of the system described in \cref{subsec:vector}. The vector $\mathbf{v}$ starts pointing along $(1,0,0)$ and rotates around axis $\mathbf{a}_1$. At one point throughout its dynamic evolution, labelled as $\mathbf{v}'$, it `jumps' state and starts rotating around $\mathbf{a}_2$. Figure made with GeoGebra \cite{hohenwarter2002, geogebra5}.}
	\label{fig:tetrahedron}
	\end{center}
\end{figure}

The second test concerns \cref{vector_path}. We now simulate the precession of a single vector of length 1 around four different axes; which axis it rotates around at any time depends on the state of the system, and can vary randomly. The four rotation axes are chosen to point at the four corners of a tetrahedron and be of length $\sqrt{3}$. All vectors start pointing along the $x$ axis (see \cref{fig:tetrahedron}).

If the axis can not change during the whole process, then we can expect

\begin{equation}\label{slow_vector}
\overline{v}^{(x)}_{slow}(t) = \frac{2}{3}\cos\left(\sqrt{3}t\right)+\frac{1}{3}
\end{equation}

while the y and z components average to zero.\newline
For this system we choose a different Markov process too. While most transitions are allowed with likelihood $(1-e^{-\nu \delta t})/3$, like in \cref{G0_uncorr}, one of the four axes is set to be an `absorbing state', from which further transitions are not possible. This does not alter the Markovian nature of the process, but alters the results quite significantly. Fundamentally we can then have three regimes:

\begin{enumerate}
	\item a slow regime during which each vector in the average precesses around its starting axis and never changes it, thus resulting in a signal defined by \cref{slow_vector};
	\item an intermediate regime in which vectors will jump axes for a finite amount of times before ending up in the absorbing state, and the signal gets dampened by the chaotic dephasing of the various vectors due to the random process;
	\item a fast regime in which vectors fall into the absorbing state very quickly and can be treated as having been in it from the start.
\end{enumerate}

\begin{figure*}[hbtp]
	\includegraphics[width=\textwidth]{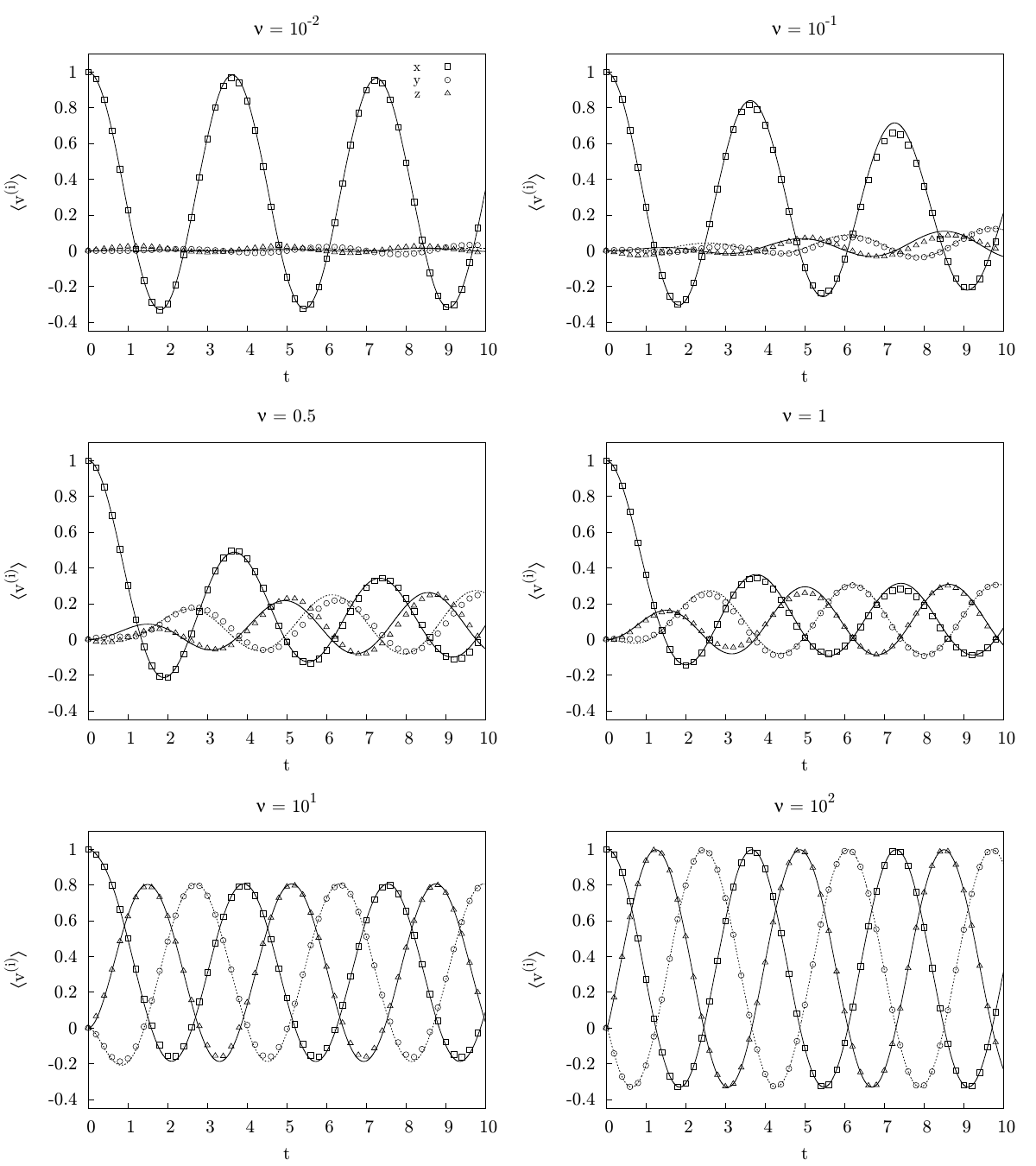}
	\caption{Average of the cartesian components for an ensemble of unit vectors starting pointing along the $x$ axis and rotating about four tetrahedral directions. The squares, circles and triangles represent respectively the $x$, $y$ and $z$ components for the Monte Carlo simulation. The corresponding lines are the Dyson equation results.}
	\label{fig:vector}
\end{figure*}

\Cref{fig:vector} illustrates these regimes pretty well.
%The average time used was 204 ms for the Dyson equation calculations and 5.62 s for the Monte Carlo ones. 
The first two frequencies illustrate a system still in the slow regime, though some decay begins appearing; the middle two are in the full intermediate regime, in which dephasing significantly affects the overall intensity of the average signal; and finally, the last two show the system reaching fully the fast regime, in which it is equivalent to a single vector rotating around a single axis.

\subsection{Density matrices}

Finally, we test the approach on a fully quantum system. Our model system in this case is a pair of coupled spin-1/2 particles subject to an exchange-like interaction as the one seen in the classic Heisenberg model \cite{mattis1981}, with the addition of a random up or down field along $z$ for each of them:

\begin{equation}\label{pair_hamiltonian}
\mathcal{H}_{\pm\pm} = -\mathbf{I}_1\mathbf{I}_2 \pm I_{1z}\pm I_{2z}
\end{equation}

\begin{figure}
 	\begin{center}
 		\includegraphics[width=0.5\textwidth]{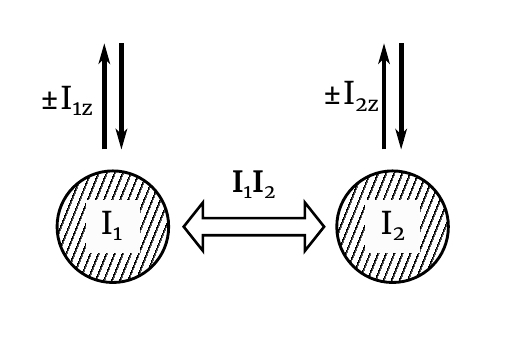}
 		\caption{Pictorial representation of the Hamiltonian in \cref{pair_hamiltonian}. Two spins are coupled by an exchange interaction, and separately they each have a Zeeman interaction pointing independently up or down.}
 		\label{fig:spins}
 	\end{center}
 \end{figure}

 (see \cref{fig:spins}). 
 
 This gives four possible Hamiltonians for the system. We consider what happens when the Hamiltonian changes in time alternating between these four possibilities - in physical terms, this could be seen as the magnetic fields applied to each spin individually flipping at random. The uncorrelated jumping matrix from \cref{G0_uncorr} was used again here. The spins are prepared in a pure state, both pointing along the $x$ axis,

\begin{equation}\label{pair_psi0}
\left|\psi_0\right> = \frac{1}{2}\left(\left|\uparrow\uparrow\right>+
\left|\uparrow\downarrow\right>+
\left|\downarrow\uparrow\right>+
\left|\downarrow\downarrow\right>\right)
\end{equation}

with a density matrix $\rho_0 = \left|\psi_0\right>\left<\psi_0\right|$. Since we are considering a system of two $S=1/2$ spins, the Hilbert space can be represented with a basis of four states. This means the density matrix and Hamiltonian can be written as 4x4 matrices. In the Fock-Liouville space, the density matrix becomes a 16 dimensional complex-valued vector, and the Liouvillian a 16x16 complex matrix that we can write as \cite{manzano2020}

\begin{equation}\label{liouvillian_mat}
\tilde{\mathcal{L}}_{\pm\pm} = i(\mathcal{H}_{\pm\pm}\otimes\mathbb{I}
-\mathbb{I}\otimes\mathcal{H}_{\pm\pm}^T)
\end{equation}

using the Kronecker product. In general, it should be noted that for a system of $n$ spins 1/2, the Liouvillian matrix is going to have $2^{4n}$ elements, and the Green's function tensors, for $N$ different possible states, will have $N^2 2^{4n}$ elements. It can be seen easily why this method, when dealing with quantum systems, can quickly scale into a very large problem, which is why for this test a relatively small and manageable system was used. A possible solution to this problem might be to discard any off-diagonal terms in the density matrix that are expected to remain very small and modify the Hamiltonian accordingly. However, in this work we will avoid further approximations and use the full density matrix for the system.\newline
%The average running times were 1.5 s for the Dyson equation calculations and 63.9 s for the Monte Carlo ones.
 Two quantities were measured: the expectation value of the correlation between the two spins, $\left<\mathbf{I}_1\mathbf{I}_2\right>$, and the Von Neumann entropy \cite{bengtsson2008}. The spins start perfectly correlated at the beginning, as they both point along $x$. We expect this correlation to oscillate even in absence of motion, as half the spins have a Hamiltonian in which the left and right Zeeman terms have opposite direction, which will cause them to dephase at some point of their evolution. However this loss of correlation should only be temporary; after a period we expect them to all return to their starting state. The correlation function should oscillate periodically between 1 and $1/2$.

\begin{figure*}[hbtp]
	\includegraphics[width=\textwidth]{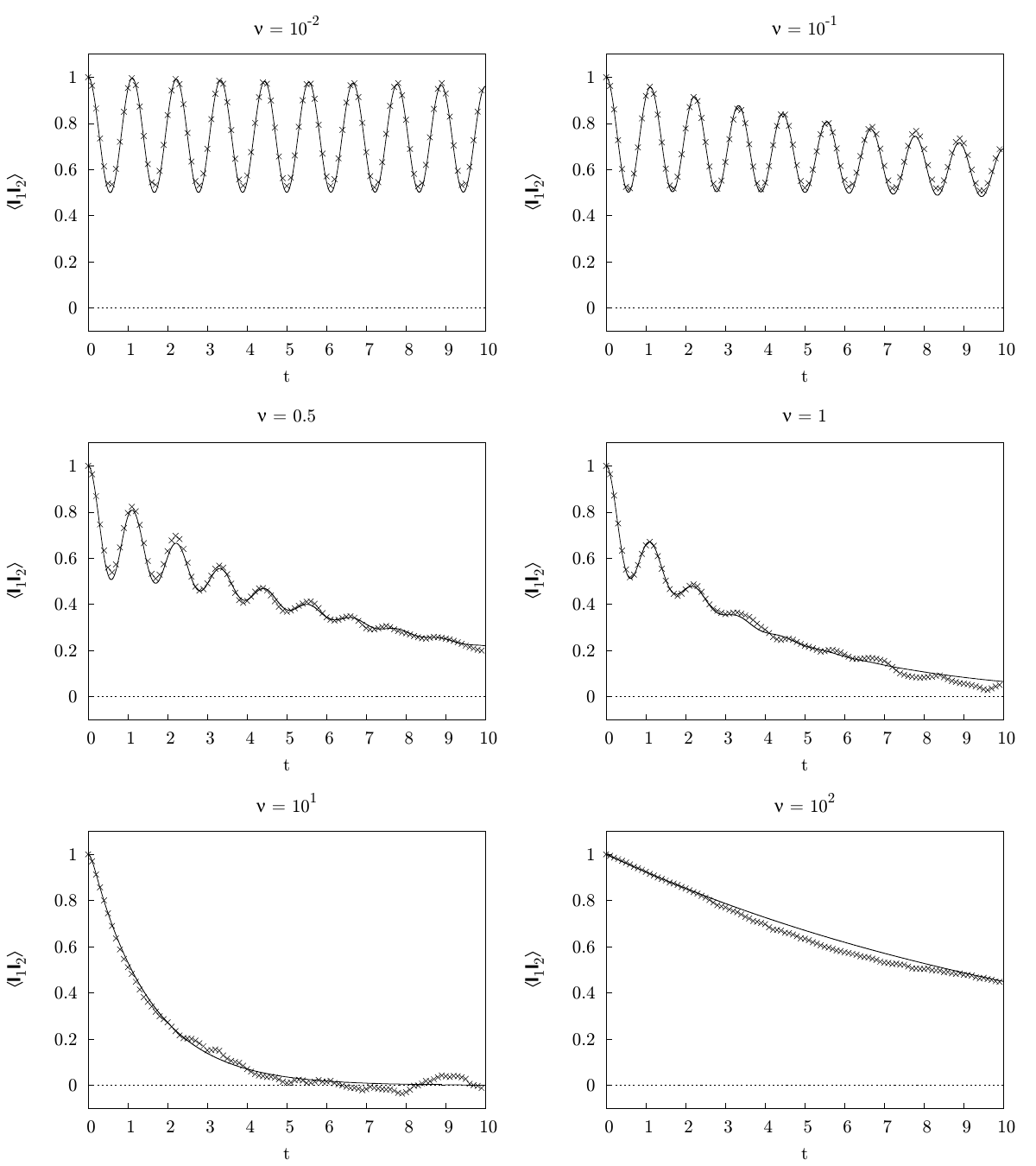}
	\caption{Expectation value of the spin pair correlation function, $\left<\mathbf{I}_1\mathbf{I}_2\right>$, for the system described in \cref{fig:spins} and different jumping frequencies. The crosses are Monte Carlo simulation results, the lines use the Dyson equation approach.}
	\label{fig:qcorr}
\end{figure*}

In \cref{fig:qcorr} we see indeed that for $\nu << 1$ this is nearly the case. In the fast limit, on the other hand, the Zeeman interaction is completely averaged out, and the spins should tend to remain perfectly correlated at all times. The intermediate regime is where we mostly expect correlation to be lost. This loss of correlation will be the consequence of irreversible loss of coherence between the various realisations of the system, which will lead to decay in the off-diagonal terms of the density matrix, and among the frequencies that were tried it shows maximally at $\nu=10$. It must also be noted that for $\nu \sim 1$ we see the biggest deviations between MC and Green's function simulations; this is likely due to the complexities of the dynamics in this regime, which mean the MC trajectory suffers the most from the effects of the relatively small sampling.

\begin{figure*}[hbtp]
	\includegraphics[width=\textwidth]{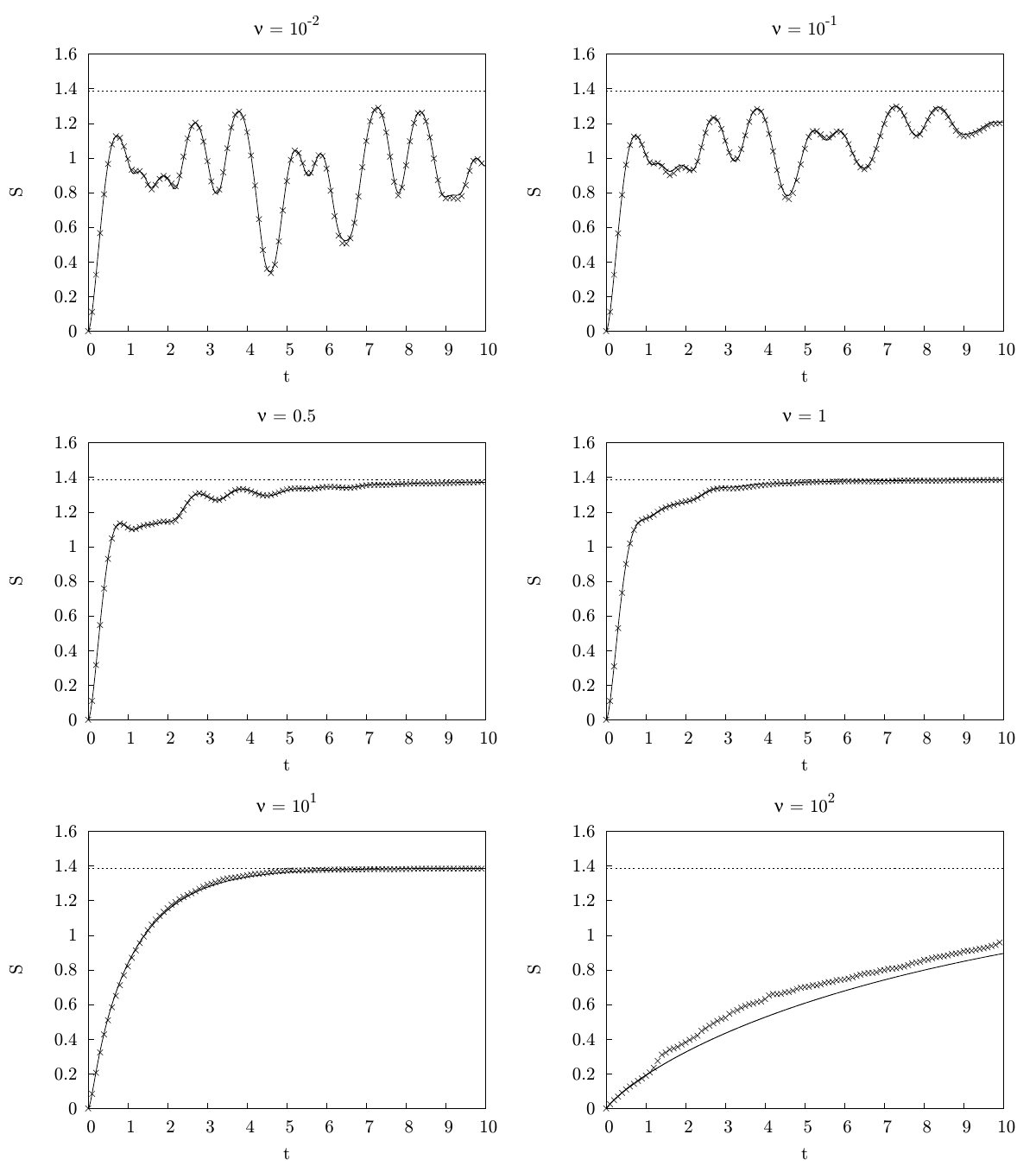}
	\caption{Evolution of the Von Neumann entropy in time for the system described in \cref{fig:spins} and different jumping frequencies. The crosses are Monte Carlo simulation results, the lines use the Dyson equation approach.}
	\label{fig:qentr}
\end{figure*}

The Von Neumann entropy, plotted in \cref{fig:qentr}, shows further evidence of this decoherence process. Since the system is initially in a pure state, it starts at 0. The maximum value, due to the size of the system, is $\log(4)$. The entropy has an oscillatory but irregular character at low frequencies, as even without motion the density matrix of the ensemble would evolve into a mixed state. However, only with higher jumping frequencies does the entropy quickly converge to its maximum value and stay there. As the frequency grows again, for $\nu >> 1$, the entropy creation process is slowed down, reflecting the averaging out of the Zeeman interaction terms which are the sources of disorder. This system therefore exhibits a nice example of decoherence, not unlike what is usually seen by applying the Lindblad master equation with some collapse operators. However, instead of using explicit interactions coupling the system with an outside environment, we implicitly represent them through random changes in the Hamiltonian.

\subsection{Performance}\label{subsec:performance}

\begin{figure*}[h]
	\begin{center}
		\includegraphics[width=0.7\textwidth]{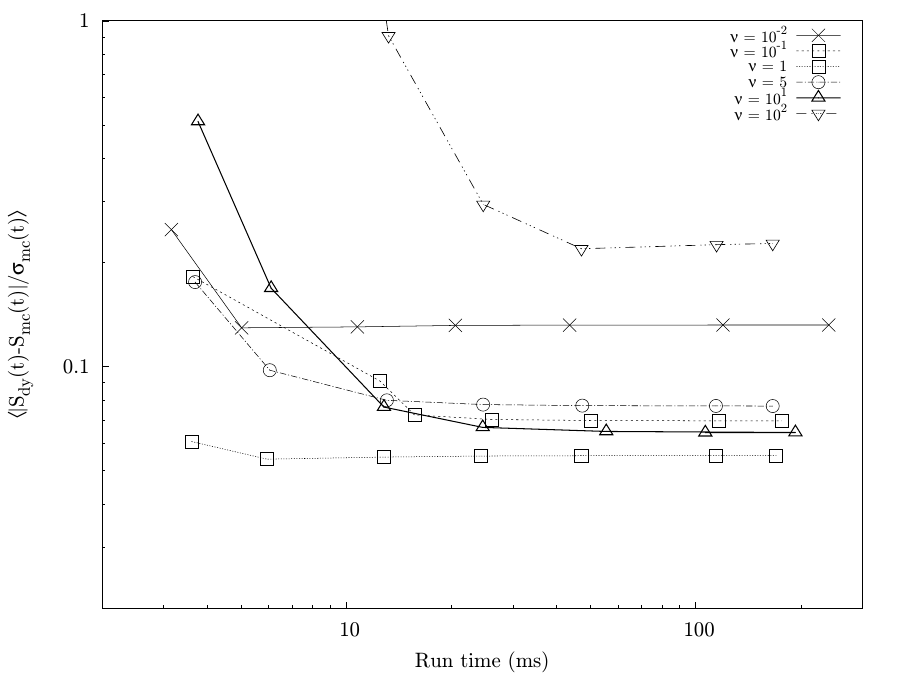}
		\caption{Runtime and mean relative error for the Dyson equation predicted evolution when solving the problem outlined in \cref{subsec:scalar} compared with a reference Monte Carlo calculation, for different time steps and jumping frequencies.}
		\label{fig:bmark}
	\end{center}
\end{figure*}

Finally, we perform a test to better characterise the accuracy of the new method. The problem chosen for the test was the one already seen in \cref{subsec:scalar}. A series of 50 Monte Carlo simulations, each with 1,000 trajectories sampled, was run on the system, for a total of 50,000 trajectories. The results were used to compute an average trajectory $S_{mc}(t)$ as well as a standard deviation $\sigma_{mc}(t)$ to characterise its error interval around the correct solution. For the Dyson equation algorithm, a number of time steps ranging from 0.1 to 0.001 were used, varying the number of steps proportionally in order to match the interval. The results were then compared with the reference solution to compute a mean relative error

\begin{equation}\label{avg_rel_err}
MRE = \frac{1}{N}\sum_{i=1}^N \frac{|S_{dy}(t_i)-S_{mc}(t_i)|}{\sigma_{mc}(t_i)},
\end{equation}

where the sum is carried over all $t_i$ for which $S_{dy}$ was computed. \Cref{fig:bmark} shows a plot of the MRE as a function of the run time of the simulation for different values of $\nu$. As it can be seen, the error quickly converges for small enough time steps (corresponding to more computationally expensive simulations) to a value on average within one standard deviation from the reference calculation. The slowest convergence is observed for $\nu >> 1$; this is likely due to the fact that for high motional frequencies the sampling needs to be far more granular in order to correctly evaluate the integral. However, this is not much of a concern, not least because in practice the fast limit for such problems as the ones solved here can be always easily retrieved analytically by simple averaging, whereas the intermediate regime, when $\nu \sim 1$, is the one that tends to be intractable by simple approximations.

\section{Conclusions}

A method previously used in an analytical solution of a specialised NMR problem in \cite{sturniolo2012} was expanded and adapted for use as a an algorithm to attack a wide class of problems. These include problems of both classical and quantum mechanics, and may be of help to develop new analysis and fitting methods in NMR and $\mu$SR in the spectral diffusion regime, in the modelling of open quantum systems and decoherence, and in the treatment of quantum field theoretical problems with approaches such as world line Monte Carlo \cite{gies2002,savkli1999} . The method was already used for educational purposes in developing the ``Muon Playground" app \cite{muonplayground}, an interactive website allowing learners to experiment with the outcome of $\mu$SR experiments in the presence of a variety of distributions of magnetic fields and dynamical regimes.

\section{Acknowledgements}

Thanks to Peter Baker for the discussions and support in developing and deploying the Muon Playground website. This work was supported by the CCP for NMR crystallography, which is funded by the EPSRC grants EP/J010510/1 and EP/M022501/1.

\bibliographystyle{ieeetr}
\bibliography{dyson_fid.bib}

\end{document}